\documentstyle[iopconf1,psfig]{article}
\begin{document}

\title{Towards an accurate determination of the age of the Universe}

\author{Raul Jimenez$\dag$\footnote {E-mail: raul@roe.ac.uk}}

\affil{$\dag$\ Institute for Astronomy, University of Edinburgh, Blackford
Hill, Edinburgh, EH9 3HJ, UK}

\beginabstract 

In the past 40 years a considerable effort has been focused in determining the
age of the Universe at zero redshift using several stellar clocks. In this
review I will describe the best theoretical methods to determine the age of
the oldest Galactic Globular Clusters (GC). I will also argue that a more
accurate age determination may come from passively evolving high-redshift
ellipticals. In particular, I will review two new methods to determine the age
of GC. These two methods are more accurate than the classical isochrone
fitting technique.  The first method is based on the morphology of the
horizontal branch and is independent of the distance modulus of the globular
cluster. The second method uses a careful binning of the stellar luminosity
function which determines simultaneously the distance and age of the GC. It is
found that the oldest GCs have an age of $13.5 \pm 2$ Gyr. The absolute
minimum age for the oldest GCs is 10.5 Gyr and the maximum is 16.0 Gyr (with
99\% confidence).  Therefore, an Einstein-De Sitter Universe ($\Omega=1$) is
not totally ruled out if the Hubble constant is about $65 \pm 10$ km s$^{-1}$
Mpc$^{-1}$.  On the other hand, the newly discovered red elliptical 53W069
($z=1.43$) provides an stronger constraint since its minimun age is 3.2 Gyr,
thus ruling out an Einstein-De Sitter Universe unless the Hubble constant is
$\leq 45 \pm 10$ km s$^{-1}$. Using 53W069 we find an age at $z=0$ of $13 \pm
2$ Gyr, in excellent agreement with the GC determination.

\endabstract

\section{Globular cluster ages}

GCs are the best stellar clocks to establish a lower limit to the age of the
Universe at $z=0$ since they fulfill the following properties: all the stars
were born at the same time, the population is chemically homogeneous and there
has been no further episodes of star formation that gave birth to new
stars which could cover up the {\it oldest} population. Support for GCs being
old comes from two facts: their metallicity is as low as 1/100 of solar and
the characteristics of their colour-magnitude diagram are those corresponding
to ages larger than 10 Gyr, i.e. the stars around the main sequence turn-off
have masses lower than 1 M$_{\odot}$. Despite the continuous effort carried
out during more than 40 years to give a precise value for the age of GCs, the
uncertainty in their age still remains about 4Gyr. The problem is
particularly complicated because age and distance have the same effect on the
morphology of the main sequence turn off point (MSTO). Deficiencies in the
input physics combined with the uncertainties in cluster distances and
interstellar reddening have made it difficult to determine globular cluster
ages with an accuracy better than about 25\%.

In this review I present two alternative methods used to derive ages of GCs
that are independent of the traditional MSTO fitting. The first method is
based on the morphology of the horizontal branch (HB), and uses {\it the
reddest points in the HB} to determine the mass of the stars in the RGB
(Jimenez et al. 1996). This method is independent of the distance modulus,
and is therefore a useful tool to compute systematics in age determinations
using the traditional MSTO fitting technique. The second of the methods is
based on a careful binning of the luminosity function (LF), and determines
simultaneously the age and distance of a GC. It is therefore a very useful
technique to determine distances to GC independently of the subdwarf fitting
and RR-Lyrae techniques. It is also very powerful in determining relative
ages of GCs with little error.

\subsection{The Isochrone fitting method}
The first (and more obvious) method to compute the age of a GC is to exploit
the fact that the locus of the MSTO in the plane $T_{\rm eff}$ vs. $L$ changes
with age (mass). In this way one computes different isochrones, i.e. tracks in
the plane $T_{\rm eff}$ vs. $L$ at the same time for all masses, with the
chemical composition of the GC to find the better fit to the MSTO region. To
do this a very important step is needed: the {\it distance} to the GC has to
be known in order to transform the theoretical luminosity into observed
magnitudes in different bands, and here is where the trouble starts. If the
distance to the GC is unknown, there is a degeneracy between age and distance
and we can simulate a different age by simply putting the GC closer or more
distant to us. If we assume the GC is closer than it really is, we will
overestimate its age.

Distances to GCs are very poorly known since it is impossible to get the
parallax of individual stars and therefore ages of GCs are not accurately
known using the isochrone fitting method. Usually, there are different methods
to compute distances to GCs: the RR-Lyrae method; the subdwarf fitting method;
the tip of the red giant branch; and the luminosity function. The RR-Lyrae
method consists in using the known Period-Luminosity relation for the RR-Lyrae
pulsators in the HB. This method gives an uncertainty of 0.25 mag in the
distance modulus determination, which translates to a 3 Gyr error in the age
determination. The subdwarf method uses the nearby low metal subdwarfs to
calibrate the distance of GCs; again its uncertainty is about 0.2 mag. The tip
of the RGB method uses the fact that stars at the tip of the RGB flash have a
well defined luminosity (Jimenez et al. 1996); therefore the tip of the RGB is
well defined and can be used as a distance indicator. The luminosity function
method is explained later in this review. It gives more precise distance
determination, and the error in the distance is only 0.05 mag. Recently, new
parallaxes of local subdwarfs determined by Hipparcos have increased the
distance inferred by the subdwarf method to GCs and have therefore reduced the
ages of GCs (e.g. Chaboyer et al. 1998)

In order to circumvent the need for the distance determinations in computing
the age, Iben \& Renzini (1984) proposed an alternative method for deriving
ages using the MSTO, the so-called $\Delta V$ method.  The method exploits the
fact that the luminosity of the MSTO changes with mass (age) and not only its
$T_{\rm eff}$. Also, the luminosity of the (HB) does not change since the core
mass of the He nucleus is the same independently of the total mass of the
stars (provided we are in the low mass range), and the luminosity in the HB is
provided by the He core burning.  Since the method is based on a relative
measure (the distance between the HB and the MSTO), it is distance
independent. Of course, the method needs the knowledge of at least one GC
distance in order to be zero calibrated. Unfortunately, the method has a
serious disadvantage: the need to know {\it accurately} the location of the
MSTO point. This turns out to be fatal for the method since it has associated
an error of 3 Gyr in the age determination.

Furthermore, all the above methods are affected by three main diseases: the
calibration colour-$T_{\rm eff}$; the calibration of the mixing-length
parameter ($\alpha$); and the need to fit morphological features in the CMD
(i.e.  the MSTO). See Table 1 for a detailed review of all errors involved in
the different methods.

The most common ages obtained for the oldest GCs using the MSTO method are in
the range 14-18 Gyr. Nevertheless, an error bar of 3 Gyr is associated with
all age determinations using the MSTO methods described above.
                                                                               
\subsection{The Horizontal Branch Morphology Method}

The spread of stars along the HB is mainly due to previous mass loss which
varies stochastically from one star to another (Rood 1973). The range of
colours where zero-age HB stars are found is a function of metallicity (the
``first parameter'') and of the range of ZAHB masses. More precisely, the ZAHB
colour at given metallicity depends on both the star's total mass and the
ratio of core mass to total mass, but the core mass is essentially fixed by
the physics of the helium flash and is quite insensitive to the mass and
metallicity. For a given average mass loss, the average final mass is thus a
decreasing function of age, which is therefore a popular candidate for the
``second parameter'' (Searle \& Zinn 1978), although other candidates such as
CNO abundance have also been suggested. A strong case for age as the chief
(though, perhaps, not necessarily the only) second parameter has been made by
Lee, Demarque \& Zinn (1994), who find a tendency for the clusters to be
younger in the outer Galactic halo. J{\o}rgensen \& Thejll (1993), using
analytical fits to a variety of RGB models and following evolution along the
RGB with mass loss treated by Reimer's (1975) formula, showed that, for
clusters with narrow RGBs (the majority), star-to-star variations in initial
mass, metallicity or mixing-length parameter can be ruled out as a source of
the spread along the HB, leaving as the only likely alternative variations in
the Reimers efficiency parameter $\eta$ (or some equivalent).

A method that is independent of the {\it distance modulus} can be developed
using the fact that the spread of stars along the HB is mainly due to previous
mass loss which varies stochastically from one star to another.  It is
therefore meaningful to proceed to an analysis of both the RGT and the HB and
to link them together to deduce general properties from morphological
arguments. The procedure that we use to analyse the morphology of the RGB and
the HB together and constrain the mass of the stars at the RGB is as follows:
Since the vertical position of the RGB depends only on metallicity and
$\alpha$, once the metallicity is known $\alpha$ is the only free parameter.
Therefore, we can find a fit for the best value of $\alpha$, using the
vertical position of the RGB. The reddest point of the HB corresponds to zero
mass loss and therefore to the most massive stars that are alive in the GC and
therefore the oldest. Using HB theoretical models we can determine the mass of
the reddest point of the HB. So it is possible to compute stellar tracks for a
certain input mass and iterate until the track at the zero age horizontal
branch matches the reddest point of the observed HB. In Jimenez et al. (1996)
we analysed eight GCs using the above method and found that the oldest GCs
were not older than {\bf 14 Gyr}.

\subsection{The Luminosity Function Method}

The LF seems to be the most natural observable to try to constrain both age
(Paczynski (1984); Ratcliff (1987)) and distance at the same time.  The LF is
a natural clock because the number of stars in a given luminosity bin
decreases with time, since more massive stars evolve more rapidly than less
massive ones. The fact that small differences in stellar masses correspond to
large differences in evolutionary time explains the power of the LF clock,
rather than being a source of uncertainty in obtaining GC ages (as it is in
the MSTO method). The LF is also a natural distance indicator, because the
number of stars in a given luminosity bin depends on the position of the bin.

For a determination of both distance and age, one needs to obtain from the LF
at least two independent constraints, which means three bins in the LF, since
one is required for the normalization. A fourth bin is also very useful in
order to check for the completeness of the stellar counts. The
second bin, the main constraint for the distance modulus, is positioned
between the RGB and the SGB (sub-giant branch), in order to partially contain
the steepest section of the LF (this gives the sensitivity to a translation in
magnitude.). The third bin, the main constraint for the age, contains the SGB,
because this is the part of the LF that is most sensitive to age.  The fourth
bin is just next to the third one, and will typically include the upper part
of the main sequence.  The procedure to obtain the LF from evolutionary
stellar tracks is illustrated in Jimenez and Padoan (1996). A power law
stellar mass function is assumed here, as in that work.

\begin{figure}
\centerline{
\psfig{figure=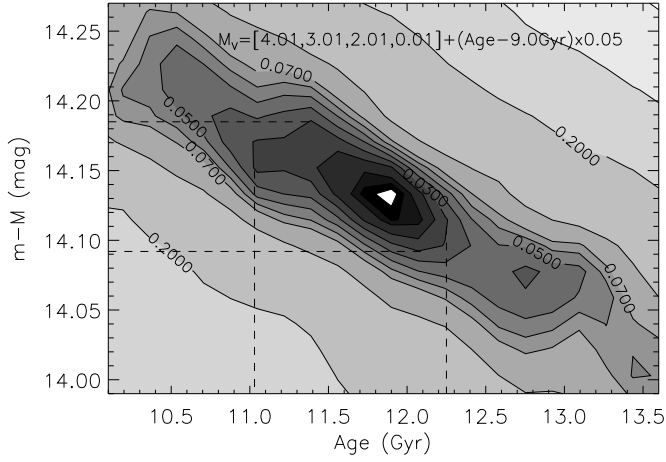,height=6cm,angle=0}}
\caption{Distance and age for the low metallicity globular cluster M55
{\bf simultaneously} determined using the LF method (see main text). The
contour lines show that the age can be determined with 1 Gyr accuracy and the
distance modulus with 0.05 provided the number of stars in the LF can be
complete to a level of 5\%.}
\end{figure}

In Fig.~1 I show the result of applying the LF method to the galactic globular
cluster M55. The plot shows contour plots for the error in the determination of
the distance modulus and age of M55 simultaneously. The contour plots
correspond to different values for the uncertainty in the number of stars in
the luminosity function. If stellar counts are within an uncertainty of 5\%,
then the age is determined with an uncertainty of $0.5$ Gyr, and the distance
modulus with an uncertainty of $0.06$ mag.  The LF method is therefore an
{\it excellent} clock for relative ages of GCs, and also a very good distance
indicator. In other words, its application provides very strong
constraints for the theory of the formation of the Galaxy.

{\it An age of 12 Gyr obtaind for M55 confirms the conclusion of the HB
morphology method that GCs are not older than 13 Gyr.}

\subsection{Discussion}

Table 1 lists the uncertainties involved in each of the three methods
described above to determine GC ages. As already discussed the MSTO method is
largely affected by the uncertainty in distance, but also uncertainties in the
mixing length, diffusion of heavy elements and in the colour-T$_{\rm eff}$
relation.

\begin{table}
\begin{center}
\footnotesize\rm
\caption{Errors associated with the different methods
described in the text to compute the age of the oldest globular clusters. The
first column lists the main uncertainties when computing GC ages.} 
\begin{tabular}{llll}
\topline
 & MSTO & HB & LF \\
 &&& \\
\midline
Distance Modulus & 25\% & 0\% & 3\% \\
Mixing Length & 10\% & 5\% & 0\% \\
Colour-$T_{\rm eff}$ & 5\% & 5\% & 0\% \\
Heavy Elements Diffusion & 7\% & 2\% & 7\% \\
$\alpha$-elements & 10\% & 5\% & 10\% \\
Reddening & 5\% & 10\% & 0\% \\
\bottomline
\end{tabular} 
\end{center}
\end{table}

The HB method uses the fact that mass loss along the RGB is the chief cause of
the HB morphology. This may seem to introduce an additional uncertainty in the
method since mass loss in low mass stars is unknown. In fact the only stars
used to determine the mass at the RGB are the reddest ones, that do not suffer
any mass loss. The evolution with mass loss along the RGB was done using the
method developed in Jimenez et al. (1996) that describes the mass loss
efficiency parameter with a realistic distribution function and minimizes its
model dependence. Furthermore, the HB method is insensitive to changes in CNO
abundances. The reason for this is that if CNO is enhanced with respect to
iron the HB becomes redder leading to a smaller mass for the reddest point of
the HB but since the stellar clock also goes faster, both effects
compensate. The HB morphology method is weakly sensitive to diffusion by heavy
elements (J. MacDonald private communication).

The LF method needs to know the metallicity of the GC. Apart from this, the LF
method is the one with the smallest errors among the three methods described
here. The biggest advantage of the LF method is that it is insensitive to the
mixing-length, redenning and colour-T$_{\rm eff}$ transformation.  Since the
LF method is based on counting stars in several bins, it is independent of
fitting to morphological features in the observed CMD of the GC. Therefore the
LF method is a superb technique to determine relative ages of GCs.

\section{Determining the age of the Universe from its high-redshift galaxies}

Traditionally, the aim of determining the age of the Universe has been focused
on stellar objects at $z=0$. The reasons are obvious: individual stars can be
resolved in stellar populations and 4m class telescopes can be used to achieve
high signal-to-noise ratios. Furthermore, at redshift larger than 0.5 is
impossible to identify the globular clusters of a host galaxy. One would wish
to identify galaxies that are themselves good stellar clocks, i.e. their
stellar population was born at once with no further episodes of star
formation. However, the most problematic issue is how to find galaxies at high
redshift. The study of `normal' star-forming galaxies at $z > 2$ has developed
into a booming astronomical industry over the past 3 years (e.g. Steidel et
al. 1997). Since most high-redshift galaxies are optically selected, one is
biased towards blue objects that show recent star formation, i.e.  biased
towards composite stellar populations with several episodes of star
formation. In this way, one can only hope to determine the age of the {\it
youngest} stars at high-redshift, not a very useful age indicator.
Nevertheless, a few valiant efforts have been carried out in order to
determine the age of the stellar populations in high-redshift galaxies
(Chambers \& Charlot 1990).

\begin{table}
\begin{center}
\footnotesize\rm
\caption{A comparison of age estimates for the stellar population of 53W069 as
derived from the instantaneous burst models of Bruzual \& Charlot (B\&C),
Worthey (1998) (W) and Jimenez et al (1999) (J99), when used to fit different
spectral indicators of age.}
\begin{tabular}{llll}
\topline
{\bf Feature} & {\bf B\&C} & {\bf W98} & {\bf J99} \\
\midline
UV-SED & 3.3 Gyr   & 3.2 Gyr & 4.0 Gyr \\
$R-K$ & 1.6 Gyr    & 3.0 Gyr & 4.0 Gyr \\
2649 \AA & 5.0 Gyr & 4.0 Gyr & 4.0 Gyr \\
2900 \AA & 4.5 Gyr & 4.2 Gyr & 4.5 Gyr \\
\bottomline
\end{tabular} 
\end{center}
\end{table}

A more promising way to find passively evolving objects is utilizing radio
galaxies. One of the cleanest results in extra-galactic astronomy is that all
powerful ($P > 10^{24}$ WHz$^{-1}$sr$^{-1}$) radio sources in the present-day
universe are hosted by giant ellipticals. It is then reasonable to assume that
high-redshift radio sources also reside in ellipticals or their
progenitors. By selecting radio galaxies at mJy flux levels we (Dunlop et
al. 1996, Spinrad et al. 1997, Dunlop 1998, Dey et al. 1999) have shown that
that it is possible to find examples of well evolved galaxies at $z \sim 1.5$
whose near-ultraviolet spectrum is uncontaminated by a recent burst of star
formation or by an AGN. Keck spectroscopy of these objects has yielded the
first detection of stellar absorption features from {\it old} stars at $z \leq
1.5$ and thus the first {\it reliable} age-dating of high-redshift
objects. The best example in our sample is 53W069 and I will concentrate in
the rest on this review in giving a robust estimate of this object's age.

\subsection{The age of 53W069}

\begin{figure}
\vspace*{-4.0cm}
\hspace*{0.5cm}
\centerline{
\psfig{figure=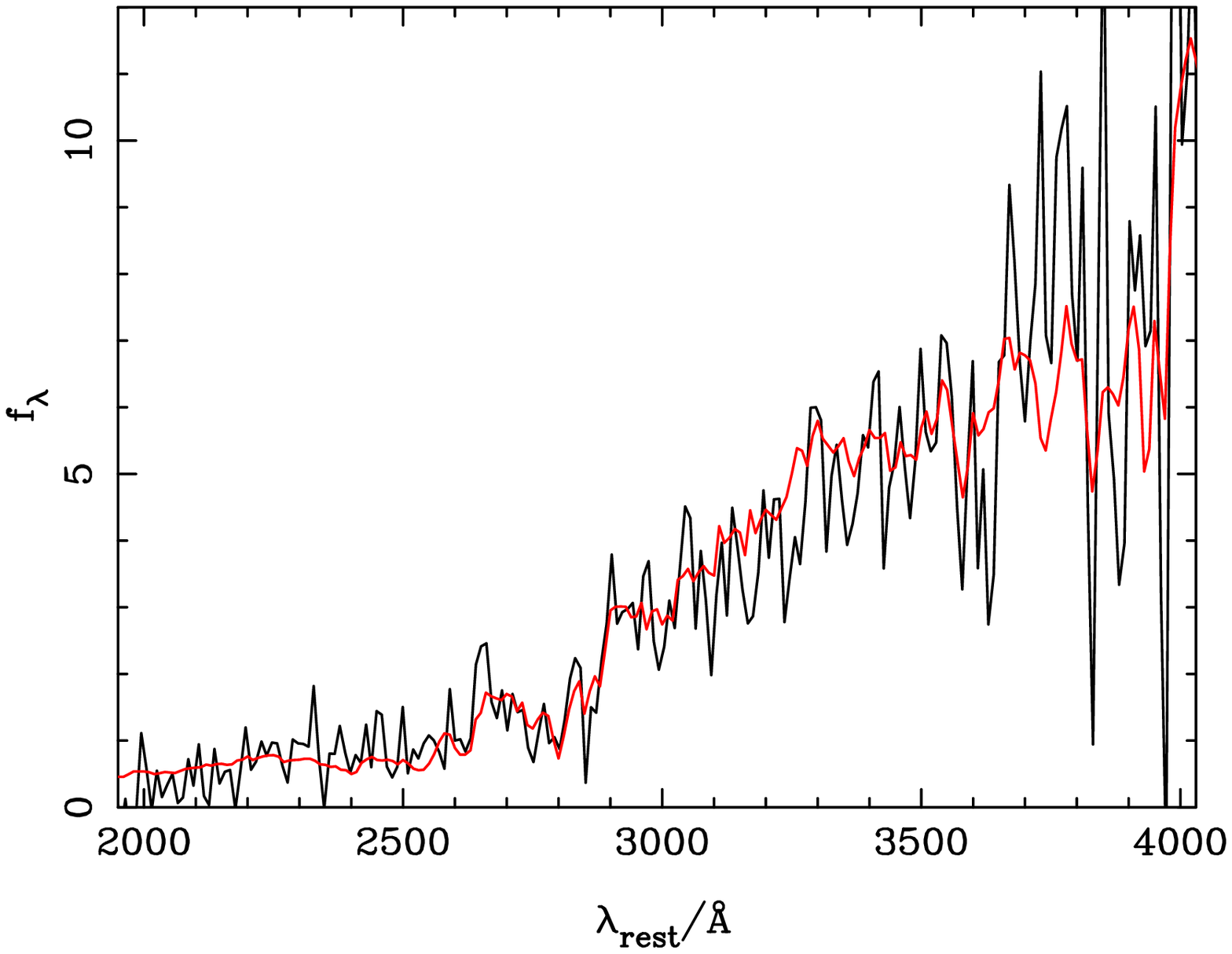,height=9cm,angle=0}
\hspace*{-1.3cm}
\psfig{figure=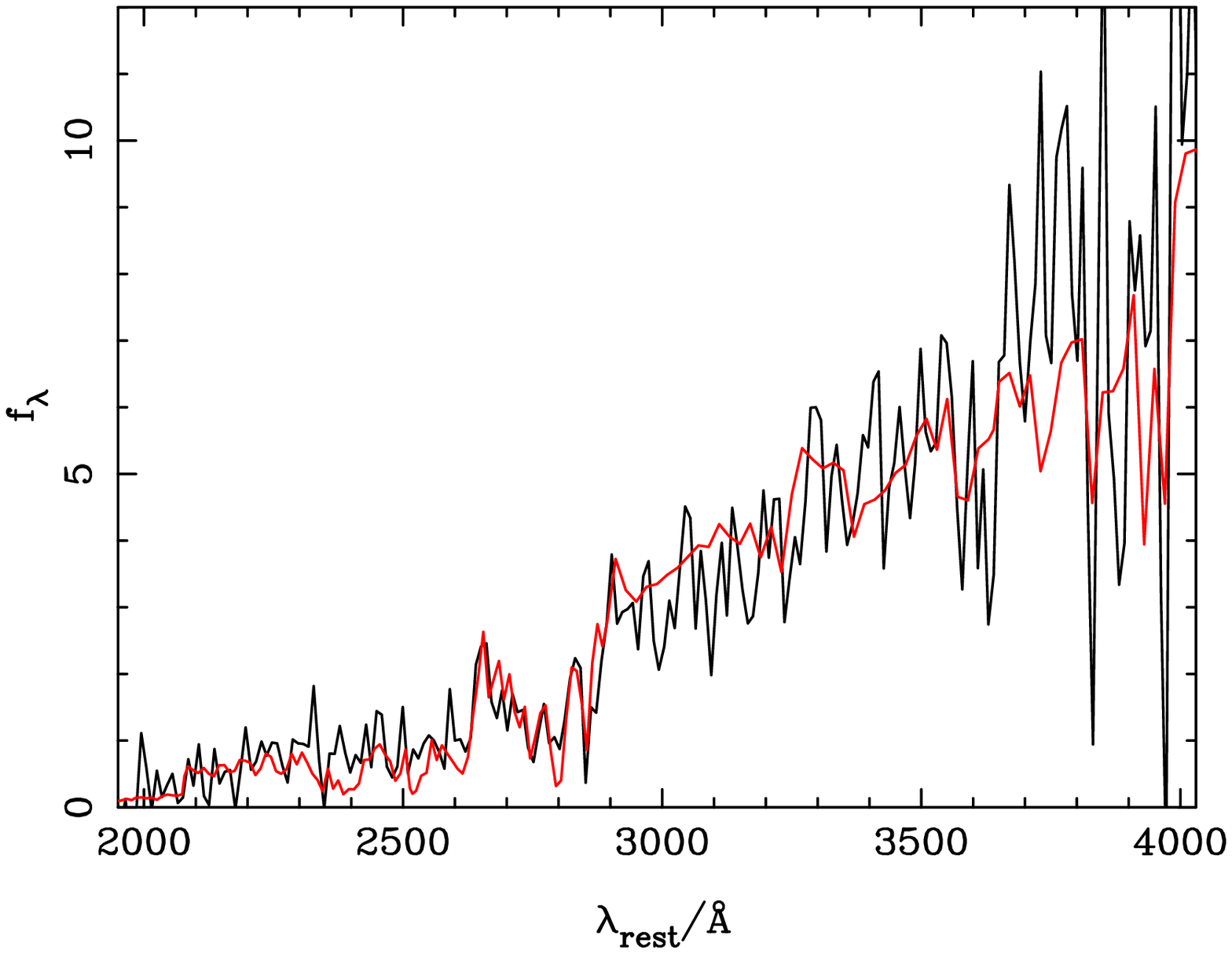,height=9cm,angle=0}}
\caption{Left panel: the spectrum of 53W069 overlaid with the best fitting
Bruzual-Charlot model, which has an age of 3.3 Gyr. Right panel: Same as
before but now overlaid with the best fitting model from Jimenez et
al. (1999), which has an age of 4.0 Gyr.}
\end{figure}

The spectral energy distribution (SED) of 53W069 is presented in Fig.~2. In
order to determine its age I have performed properly weighted chi-squared fits
to the ultra-violet SED using the popular Worthey and also Bruzual \& Charlot
synthetic stellar population models and the models developed by our group
(Jimenez et al. 1999). The results are listed in Table 2. A few important
points are obvious from this table. First, all models yield ages larger than 3
Gyr for 53W069. Second, column 1 shows that the Bruzual \& Charlot models seem
to be internally inconsistent in the sense that they are capable of
reproducing very red $R-K$ colours at a much younger age than they can
reproduce the ultraviolet SED or the spectral breaks. However, since $R-K$ is
mainly affected by the evolution of the late stages of stellar evolution, once
should focus on ages derived from the UV-spectrum since this only depends on
the correct prediction of the MSTO, a simpler and thus an easier part to
model. Indeed, not do so it tantamount to throwing away the new, more robust
information which can be gleaned from the spectroscopy. If one focuses on the
results of 53W069 then, ignoring the anomalously young $R-K$ age produced by
the Bruzual \&Charlot models, all models are basically in good agreement that
the overall shape of the UV SED and the strengths of the main spectral
features are consistent with an age in the range 3.0 to 4.0 Gyr, and certainly
yielding a robust (99\% confidence) minimum age of 3.0 Gyr.

Spectroscopically, 53W069 thus appears to be the best known example of old,
passively-evolving elliptical galaxies at redshifts as high as $z \sim 1.5$.
Using WPFC2 and NICMOS images below and above the 4000 \AA \, break we have
verified that this also holds for its morphological properties and
scalelengths (Dunlop 1998). Using a 2-dimensional fitting code it is possible
to show that 53W069 is consistent with a $r^{1/4}$ law and inconsistent with a
exponential disc profile. Furthermore, a physical half-light radius of $r_{e}
\approx 4$ kpc has been obtained assuming $\Omega=1$ and $H_0 = 50$ km
s$^{-1}$ Mpc$^{-1}$, which lies exactly in the Kormendy relation for
ellipticals.

In summary, the new data on 53W069 clearly show that the Universe at $z \sim
1.5$ contains stellar systems whose populations are 3 to 4 Gyr old. At $z \sim
1.5$ the Universe was less than 30\% of its present age, and the uncertainties
are largely independent of those encountered in GCs studies. The existence of
53W069 permits only low Hubble constants and/or low cosmic densities; in
particular, an $\Omega=1$ Universe requires $H_0 \leq 45$ km s$^{-1}$
Mpc$^{-1}$ . On the other hand, in an Universe with a cosmological constant,
53W069 requires $\Omega_{\lambda} \geq 0.4$ if $H_0 \geq 60$ km s$^{-1}$
Mpc$^{-1}$

\section{Conclusions}

The main conclusions of this review are:

\begin{enumerate}

\item The two methods presented in this paper agree on an age of about 13
Gyr for the oldest GCs. The minimum possible age is 10.5 Gyr and the maximum
16 Gyr, with 99\% confidence.

\item A more accurate determination of the age of the Universe can be obtained
using high-redshift elliptical galaxies. 53W069 is one of the reddest objects
at $z \sim 1.5$ and has an age of at least 3 Gyr. This yields to an age of
the Universe of 13 Gyr at $z=0$, in excellent agreement with the GC
determination. This age totally rules out an Einstein de-Sitter Universe
unless $H_0 \leq 45$ km s$^{-1}$ Mpc$^{-1}$.

\end{enumerate}

\section*{Acknowledgments}

I am grateful to Dave Bowen for a careful reading of the manuscript.  This
paper draws on unpublished collaborative work with James Dunlop, John Peacock,
Arjun Dey \& Hy Spinrad.


\begin{references}
\item{} Chaboyer, B., Demarque, P., Kernan, P.J., Krauss, L.M., 1998, ApJ,
494, 96.
\item{} Chambers, K.C., Charlot, S., 1990, ApJ, 348, L1.
\item{} Dey et al., 1999, in preparation
\item{} Dunlop, J., 1998, astro-ph/9801114.
\item{} Dunlop J., Peacock J., Spinrad H., Dey A., Jimenez R., Stern D., Windhorst R., 1996, Nature, 381, 581.
\item{} Iben I., Renzini A. (1984) Phys. Rep., 105, 329.
\item{} Jimenez et al., 1999, MNRAS, in press.
\item{} Jimenez R., Padoan P. (1996) ApJ 463, 17L.
\item{} Jimenez R., Thejll P., J{\o}rgensen U.G., MacDonald J., 
Pagel B. (1996) MNRAS, 282, 926.
\item{} J{\o}rgensen U.G., Thejll P. (1993) A\&A, 272, 255.
\item{} Lee Y., Demarque P., Zinn R. (1994) ApJ, 423, 380. 
\item{} Ratcliff S. (1987) ApJ, 318, 196.
\item{} Reimers D. (1975) `Circumstellar envelopes and mass loss 
of red giants', In: Problems in stellar atmospheres and envelopes, 
Springer-Verlag, p. 229. 
\item{} Rood R.T. (1973) ApJ, 184, 815.
\item{} Searle L., Zinn R. (1978) ApJ, 225, 357.
\item{} Spinrad H., Dey A., Stern D., Dunlop J., Peacock J., Jimenez R., Windhorst R., 1997, ApJ, 484, 581.
\item{} Steidel et al., 1997, astro-ph/9708125.
\item{} Paczynski B. (1984) ApJ, 284, 670.
\item{} Padoan P., Jimenez R. (1997) ApJ 475, 580.

\end{references}
\end{document}